# The Evolution of Keylogger Technologies: A Survey from Historical Origins to Emerging Opportunities

Marco Salas-Nino[1], Grant Ritter[1], Daniel Hamdan[1], Tao Wang[2], and Tao Hou[1]

[1]Texas State University, {mas740, hvx1, orw17, taohou}@txstate.edu

[2]University of North Carolina at Charlotte, twang27@charlotte.edu

## Abstract

As the digital world evolves, so do the threats to our security do too. Keyloggers were once a large threat to the cyber world. Though undergoing many transformations alongside the technological advancements of today, it is important to raise questions about the importance of Anti-Keyloggers in our current state of cyber security. This survey dives into the historical evolution of Keyloggers and investigates their current day forms. Within this inspection of Keyloggers, we must propose whether Anti-Keyloggers serve a purpose to this ever-changing landscape before us or if emerging strategies have rendered them obsolete.

## 1 Introduction

It is well known that humans are the weakest link in cybersecurity. With users passing through multiple sites a session, they are bound to pick up unwanted malware/spyware by mistake which can lead to all sorts of information being stolen. One of these software's is called a Keylogger. Keyloggers stay hidden on a victim's computer stealing their keyboard's inputs. This data can be used to find valuable information and passwords. With the internet becoming increasingly important for day-to-day use, the security behind it is ever increasing. As of 2023, we have reached leaps and bounds to strive to protect our information to the best of our ability and that leaves us with our two questions: do Keyloggers pose a threat anymore? More importantly, do Anti-Keyloggers serve a purpose to us in the future?

## 2 Keylogger Technologies

Keyloggers are tools used to record keystrokes and user activity on a computer [2, 11]. One of the fundamental ways a keylogger records a user's activity is recording everything a user types, however there are a lot of distinct types of keyloggers that record different things. There are software and hardware keyloggers. Whichever keylogger an individual has they all aim for the same objective. Recording what the user typed or did on his or her computer and report the information back to the attacker. Our goal here is to bring awareness of keyloggers and understand the dangers a keylogger can bring.

*2.1 Hardware Keyloggers*

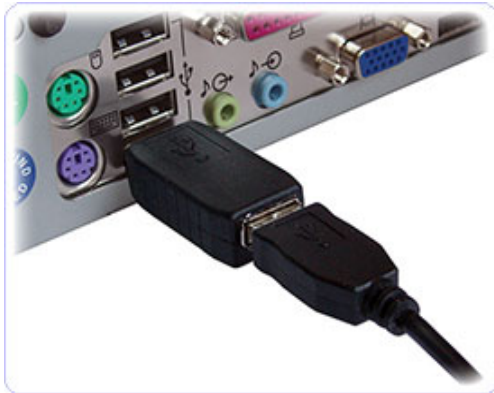

**Figure 1** Hardware Keylogger USB Connector

A regular hardware keylogger logs keystrokes using a hardware circuit that attaches between the computer and the keyboard [9]. All information that attempts to flow from the keyboard to the computer passes through the hardware, allowing the information to be stored in internal memory. The advantages of a hardware logger come from its ability to run independently from the computer and its operating system. This allows for the Keylogger to avoid interference with any machine software and gives anti-malware a much harder time to detect it. Firmware-based keyloggers is a BIOS-level firmware resulting from modifications to record process events. This can be physical or root-level access which is required for the machine. The software loaded into the BIOS needs to be created for the specific hardware the machine uses. Wireless keyboard/mouse sniffers are passive data collectors that try to take any data signals passed between wireless devices and the computer. To prevent this sort of hardware tampering, encryption is typically used to hide the data passed between the devices. The last trick these sniffers typically contain is the ability to enter commands into a victim's computer. This bypasses the encryption entirely due to it no longer exclusively snatching data.

*2.1.2 Software Keyloggers*

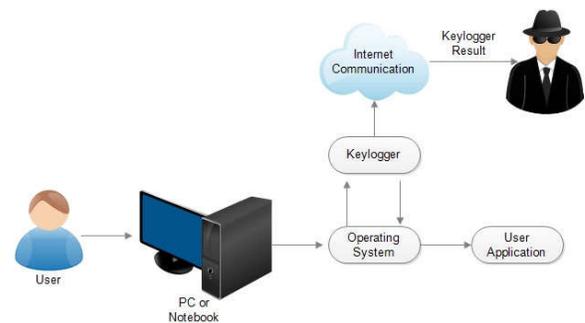

**Figure 2** Software Keylogger Path

Software keyloggers do not require any physical device to gain access to a device. Users may accidentally download this software through surfing the web. With many methods of picking up one of these software, there are different methods this software uses to gather a victim's data [33]. Form-Grabbing software Keyloggers gather data entered into a field. Rather than being directly downloaded to a victim's computer, this type of software is deployed on a website. An attacker may use a Form-Grabbing Keylogger on a malicious site which prompts the user to enter login information. JavaScript keyloggers, written in JavaScript, are injected into websites. This software records events whenever you press a key within the application.

Kernal-based Keyloggers work their way into the system's core for admin-level access. These loggers can then bypass any restrictions to access data entered into a user's system. API-based keyloggers directly take the data sent between the signals sent from each keypress to the current program. Since Application programming interfaces (APIs) allow developers and hardware manufacturers to integrate their products, API keyloggers intercept this information to log the data sent elsewhere.

*2.2 How can you accidentally obtain a Software Keylogger?*

Just like many different forms of malware/spyware, Keyloggers can be obtained by accident in similar ways [1]. Spear phishing or fake email is one of the easiest ways for hackers to deliver Keyloggers into a victim's computer. Hackers will often try to attach malware containing their methods of attack to emails and mass deliver them to random victim's hoping that any of them trigger their trap. Even the simple act of opening the email may be enough to trigger these attacks. Installing suspicious software is another way of contracting a Keylogger. By embedding Keyloggers into seemingly trustworthy software, hackers can collect the information from the victim's computer like any other Keylogger. Due to the escalating number of applications that users download to their devices; they are bound to eventually pick up some form of spyware. Phishing links are infected links that redirect you to unwanted websites. Users can encounter these URLs countless times while surfing the web. Once a victim visits one of these infected websites, a keylogger can latch onto their system within a few seconds installing the spyware. Finally, Drive by Downloads are a common way to get a Keylogger attached to your system. Like Phishing links, this can install spyware to your device. This can be done by accidentally passing by an infected site while you busy yourself by browsing the web.

## 3 Keylogger Evolution

Keyloggers are not a new form of data collection. Although Keyloggers are a much more advanced threat now, there was still a point where the idea had to come from. It is with the evolution of technology and the internet that attackers had to also evolve their attacks alongside it.

*3.1 Selectric Bug*

The first recorded keylogger was in the 1970s. The Soviet Union wanted to get information inside the U.S. embassy [32], but they had one obstacle. How can they get consistent information and not waste endless number of spies? At the time there were no computers, there were just electric typewriters. This is when the Soviet Union developed a hardware keylogging device for electric typewriters. It was called the Selectric bug. The selectric bug tracked the movements of the printhead by measuring the magnetic field released by the movements of the printhead [32]. This keylogger specifically targeted the IBM electric typewriter the ones that were used by U.S. diplomats in the U.S. embassy. In addition, it was also used to consulate buildings in St. Petersburg and Moscow. The Soviet Union spied and gathered information using this device and it worked for a few years up until 1984. Soon as they caught on, they found Selectric bugs on sixteen typewriters.

### 3.2 Perry Kivolowitz

In 1983 another keylogger was made by Perry Kivolowitz [32]. Specifically, it was a software keylogger that located and dumped character lists in a Unix kernel. As the years passed keyloggers have been growing and began to broaden around the 1990s. Since then, a lot more keylogger malware was created. In other words, attackers no longer used techniques such as stealing personal data, and credit card numbers, but also targeted home users for fraud and phishing purposes.

### 3.3 Ghost Keylogger

In 2000, there was a software known for being a cyber ghost and was given the name the ghost keylogger [18]. This was one of the earliest times where keyloggers were heard, creating a sense of fear. At the time, the internet was new and computer users did not know how to defend themselves from these attacks. This software was known for being in everyone's computer. Recording their keystrokes and information. It affected individual people, and small businesses. The attackers who made this software are unknown, but they left a big print in the digital world. Affecting individuals from different regions across the world [18]. From a Californian learning how to use the computer to a small business in Australia. It is unspecified the exact financial damage or the number of users affected by the ghost keylogger. What is known is the new form of cyber threats. It is believed the ghost keylogger is the one that started the path for new keyloggers to be created.

### 3.4 Zeus Keylogger

In 2007, there was a keylogger growing that set fear to those who heard it. They feared it so much it was giving the name Zeus due to the destruction it brought. The Zeus keylogger was used by cybercriminals in an advanced organized network [18]. When attackers used this keylogger they specifically targeted financial institutions and random people worldwide. Financial institutions like insurance companies, credit unions, brokerage firms, etc. No person was safe from this keylogger. The Zeus keylogger was responsible for major implications and damage. That was around the hundreds of millions of dollars. The people who were hurt by this were in the millions. The keylogger grabbed sensitive information such as passwords, usernames, and other sensitive banking information. This invoked feelings of distrust across the globe, the people no longer felt secure about online banking. To regain the trust of the people in cybersecurity a battle had to be fought. The battle against the Zeus keylogger is an important event in cybersecurity. Every individual in cybersecurity imaginable worked together from the international community, cybersecurity firm, all the way to the law enforcements, to stop the Zeus network from inflicting anymore damage on innocent civilians. Thanks to these collaborations it allowed arrest and convictions to the attackers at fault. Many individuals who are in the cybersecurity field still remember this major event in cybersecurity history. It remains as a reminder of the impact a keylogger has as dangerous as Zeus can do to any person around the world. Not only people but financial institutions as well.

*3.5 Nordea Bank Attack*

In 2007 a Swedish bank by the name Nordea got hit by one of the biggest online bank heists ever raking over $1.1 million [19]. Nordea customers said they were getting sent e-mails containing a tailor-made trojan. In other words, customers believed it was the bank sending them e-mails, it affected about 250 customers. It is believed organized Russian criminals orchestrated the attack. They sent an Email in which it encouraged the victims to download a suspicious application. Downloading a file raking.zip that was infected by the Trojan. Security companies call this haxdoor.ki. A Haxdoor installs keyloggers to record keystrokes, and hides using a rootkit. The keylogger was activated once the victims tried to log in to their Nordea online banking information. However, the victims were putting their log-in and sensitive information on a false home page. The victims kept receiving error messages from the false home page. Once the attackers successfully used the keylogger they took money from the victim's account. This attack lasted for 15 months, sending small transactions to themselves so overall they can have $1.1 million. The attackers used this strategy to not raise awareness to the victims getting robbe. Later, McAfee discovered that the log-in information went to servers in the US and Russia. The attackers are still yet to be found. A keylogger gave the attackers a once in a lifetime paycheck.

*3.6 GTA V Modification Attack*

Mods short for modifications are software designs that give a fresh style and feel to a computer system or in other words PC. Mods play a huge role for PC gamers; they give thrills to those who enjoy gaming in the comfort of their home in front of a computer. However even keyloggers can appear in video game mods. In 2015 a mod for the game Grand Theft Auto V was uploaded to a website called GTA5-mods.com [32, 34]. The website owners and gamers were unaware of the fact there was a hidden keylogger. It was discovered when gamers noticed suspicious behavior from the jet planes in the game. As the gamers searched their PC, they noticed an odd C# compiler program running in the background, receiving, sending, and processing information across the web. It was later found that Fade.exe executable was tracking their computer activity. Such as what the user typed and switching from one window to another. However, the GTA V mod was removed from the site, and the website owners apologized for the incident.

*3.7 HP Laptop 2017 recall*

In 2017, HP laptops had a keylogger. These HP devices logged keystrokes in a file and would send them over to a debugging API [11]. Creating a security breach since it allowed local users to be able to retrieve passwords and other sensitive data from victims. It was believed to be negligent by the developers rather than intentional.

In which they claimed it to be used as a debugging tool, thankfully though HP did a recall and fixed the situation.

## 4 Impacts of Keyloggers

Although it may appear that keyloggers only bring negatives to the world of Cyber security, there are a few ethically ambiguous uses that may be more beneficial than harmful [15, 16]. Parental supervision is important for protecting children from accessing inappropriate sites. Parents can use Keylogging software to monitor their child in their online and social activities. Next, keyloggers have been used in large companies who need to monitor substantial amounts of employees. With full transparency of the use of keylogger to your employees, keyloggers can increase productivity by cutting wasted time by monitoring wasted time. Wasted time can occur by chatting with others online or watching entertainment. Monitoring employees may reduce corruption within the company by tracking bad activity. Lastly, in case there is any legal trouble within the company, keyloggers can be used as evidence for protection. At any given time, there is stored data on the happenings in a company.

Like previously stated, Keyloggers contain a lot of downsides to cyber security [15]. Regarding privacy invasion, Keyloggers can be used to monitor someone's keystrokes without their consent which is a violation of their privacy. Keyloggers are a large contributor to Data Theft due to their ability to capture sensitive information, such as passwords, credit card numbers, and personalized messages. Consent and Transparency present ethical concerns for keyloggers that are used without the consent of the individual being monitored. Transparency allows for people to be aware their actions are being monitored. Next, unproperly handled keyloggers can become a security risk if they are vulnerable to attacks or if the captured data is not properly secured. Keyloggers have many Legal Implications, in many jurisdictions, using keyloggers without proper authorization is illegal. With their given ability to capture and store data, they Keyloggers have a large possibility of being misused for malicious intent.

## 5 Anti-Keyloggers

With such an oppressive force such as Keyloggers, we have had to produce ways to specifically target Keyloggers directly. Anti-Keyloggers are exactly as they sound as they are designed to stop Keyloggers. Anti-keyloggers are software designed in such a way that it can detect any keylogger present in the system. Specifically, it can prevent the act of Keylogging [4, 8, 23]. This protects the user from getting their keystrokes captured, so the use of an Anti-keylogger stops attackers from stealing confidential/sensitive information. It is common to see these sorts of Anti-spyware programs in public machines. The two forms of detection an Anti-Keylogger software may use are Signature-based Detection and Behavior detection. Signature-based detection checks if any of the files on the computer have a similar presence to a keylogger and marks it as a threat to the computer. The disadvantage of Signature-based detection is that it can only protect from listed keylogger activity. It does not stop unrecognized keyloggers. Behavioral Detection is also known as heuristic. It contains properties which enable them to identify keylogger activity that could be considered as harmful to the computer.

### 5.1 Advantages of Anti-Keyloggers

When discussing the relevance of Anti-Keyloggers, we need to figure out what they bring to the table [36]. Firstly, Anti-Keyloggers allow the management of activities of any Keyloggers present in the system. If it identifies said Keyloggers, then it prevents its activity from saving crucial information. Although this can be turned into a Disadvantage, it does not make an effort to identify the types of keystrokes. This means it's not able to differentiate between the legal keystroke program and illegal keystroke programs. Just like how companies use Keyloggers to monitor work done, companies can use Anti-Keyloggers to keep intruders out. Depending on which Anti-Keylogger the user uses, it can provide a seamless product that has no delay to the inputs from the keyboard. For users who code in C++, Anti-Keyloggers are easily integrated with C++ libraries.

### 5.2 Disadvantages of Anti-Keyloggers

Since most Keyloggers are unable to differentiate between a legal keystroke program and an illegal keystroke program, it may identify the legal program as a threat [36]. This can easily be countered with a whitelist mechanic that allows the users to target specific programs they wish to no longer interrupt. Unfortunately, the major downside that Anti-Keyloggers contain is that they have a hard time detecting hardware Keyloggers due to their ability to act independently of the machine.

## 6 Discussions

### 6.1 The Rates of Spyware/Keyloggers

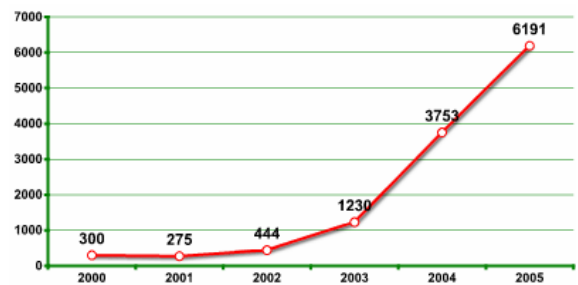

**Figure 3** Early look at Keyloggers

Looking at early charts depicting the growth of keyloggers around the early 2000's, Keyloggers seemed to be on the rise around 2005. According to the research of John Bambenek, an analyst at SANS institute [11], approximately 10 million computers in the US alone were infected with malicious programs that contain some sort of keylogging function. There was a clear interest shown by attackers that Keyloggers were a valuable asset to their operations. It would be expected that this trend would continue.

Examining this graph from 2023 [7], we can notice that there is a noticeable increase of total malicious programs and potentially unwanted applications (PUA). Assuming that Keyloggers and all their forms make up a small percentage of these graphs similar to the total shown in the early 2000's of Keyloggers. Then we can form a conclusion that Keyloggers still exhibit problems even in 2023.

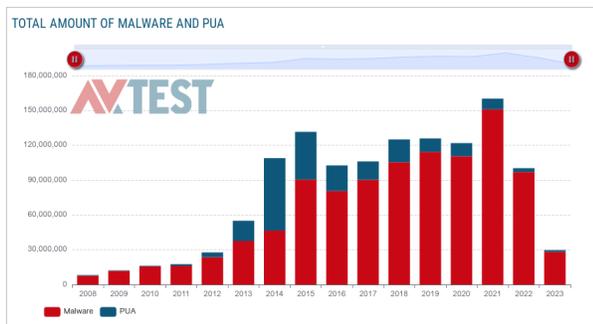

**Figure 4** Total Newly Developed Malwares

*6.2 What does this mean for Anti-Keyloggers?* By going through the definition, the brief history of Keyloggers, and the rates at which malware is created, we can determine that the continuous use of Keyloggers is still an ever-evolving threat that still needs solutions to prevent them. Although they may seem rather simple, they still propose a challenge to cyber security that must be dealt with. Anti-Keyloggers present a solution that specifically targets this issue that provides simple security to those who require it such as public equipment services, companies, gamers, and users that just want security. With the ever-evolving Keyloggers, we can expect Anti-Keyloggers to evolve alongside them just as Keyloggers have evolved around evolving machines.

# 7 Conclusions

This leaves us with the questionable future of Anti-Keyloggers. Even with technology evolving around us, Keyloggers will need to adapt to create new attacks. Currently, the latest development made in attacks has been the 'BlackMamba' keylogging attack [35]. This Keylogger is a proof-of-concept artificial intelligence – driven cyberattack that changes its code on the fly to slip past automated securities such as the current Anti-keyloggers of today. This technology is powered using a large language model which is a type of language model known for its ability to understand general-purpose languages and language generation [37]. It is up to future developers to predict these AI Keyloggers and evolve Anti-Keyloggers to combat them.